\documentclass[aps,prd,twocolumn,showpacs,superscriptaddress,groupedaddress,nofootinbib]{revtex4-2}  % Must update version
 \usepackage[portuges, english]{babel}
 \usepackage[utf8]{inputenc}
 \usepackage{amsmath}
 \usepackage[makeroom]{cancel}
 \usepackage{slashed}
 \usepackage{enumitem}
 \usepackage{times}
 \usepackage{natbib}
\usepackage[hidelinks,colorlinks=true,linkcolor=blue,citecolor=magenta]{hyperref}
\usepackage{tikz,hyperref, xcolor} 
\definecolor{lime}{HTML}{A6CE39}
\definecolor{Emerald}{HTML}{50c878}
\definecolor{PineGreen}{HTML}{01796F}
\definecolor{ForestGreen}{HTML}{228B22}
\definecolor{Coral}{HTML}{FF7F50}
\definecolor{YellowOrange}{HTML}{E94E16}  
 \DeclareRobustCommand{\orcidicon}{%
 	\begin{tikzpicture}
 		\draw[lime, fill=lime] (0,0) 
 		circle [radius=0.16] 
 		node[white] {{\fontfamily{qag}\selectfont \tiny ID}};
 		\draw[white, fill=white] (-0.0625,0.095) 
 		circle [radius=0.007];
 	\end{tikzpicture}
 	\hspace{-2mm}
 }
 \foreach \x in {A, ..., Z}{%
 	\expandafter\xdef\csname orcid\x\endcsname{\noexpand\href{https://orcid.org/\csname orcidauthor\x\endcsname}{\noexpand\orcidicon}}
 }
\def\araa{ARA\&A}             % Annual Review of Astron and Astrophys
\def\apj{ApJ}                 % Astrophysical Journal
\def\apjl{ApJ}               % Astrophysical Journal, Letters
\def\apjs{ApJS}              % Astrophysical Journal, Supplement
\def\jcap{JCAP}              % 
\def\mnras{MNRAS}             % Monthly Notices of the RAS
\def\prc{Phys.~Rev.~C}        % Physical Review C
\def\prd{Phys.~Rev.~D}        % Physical Review D
\def\prl{Phys.~Rev.~Lett.}    % Physical Review Letters
\def\nat{Nature}              % Nature
\def\physrep{Phys.~Rep.}   % Physics Reports
\begin{document}
		
%%%%%%%%%%%%%%%%%%%%%%%%%%%%%%%%%%%%%%%%%%%%%%%%%%%%%%%%%%%%%%%%%%%%%%%%%%%%%%
%\acknowledgments
%\section*{Acknowledgments}
\medskip\noindent
\title{\large Exploring nontandard quark interactions through solar neutrino studies}
		
%\author{Il\'idio Lopes}
%\email[]{ilidio.lopes@tecnico.ulisboa.pt}
%\affiliation{ Centro de Astrof\'{\i}sica e Gravita\c c\~ao - CENTRA,
%Departamento de F\'{\i}sica, Instituto Superior T\'ecnico - IST,\\ 
%Universidade de Lisboa - UL, Av. Rovisco Pais 1, 1049-001 Lisboa, Portugal \\}
%\author[0000-0002-5011-9195]{Ilídio Lopes}
\author{Ilídio Lopes\orcidA{}}
\affiliation{Centro de Astrof\'{\i}sica e Gravita\c c\~ao  - CENTRA, \\
	Departamento de F\'{\i}sica, Instituto Superior T\'ecnico - IST,\\
	Universidade de Lisboa - UL, Avenida Rovisco Pais 1, 1049-001 Lisboa, Portugal}
\email{ilidio.lopes@tecnico.ulisboa.pt}

\begin{abstract}
We investigate the effects of a nonstandard interaction  NSI) extension of the standard model of particle physics on solar neutrino flavor oscillations.  This NSI model introduces a $U_{Z^\prime}(1)$ gauge symmetry through a $Z^\prime$
boson that mixes with the photon, creating a neutral  current between active neutrinos and matter  fields via a unique coupling to up and down quarks. The interaction is defined by a 
single  parameter, $\zeta_o$, which is related to the  $Z^\prime$ boson's mass $m_{Z^\prime}$ 
and coupling constant $g_{Z^\prime}$. Notably, this model relaxes the bounds on coherent elastic neutrino-nucleus 
Scattering experiments and fits the experimental values of the anomalous magnetic dipole moment of the muon.
In this study, we use solar neutrino measurements and an up-to-date standard solar model to 
evaluate the neutrino flavor oscillations and assess the constraints on $\zeta_o$.
Our study indicates that the NSI model aligns with the current solar neutrino data when $\zeta_o$ is between $-0.7$ and $0.002$. These models have $\chi^2_{\nu}$ values equal to or better than the standard neutrino flavor oscillation model, which stands at a $\chi^2_{\nu}$ of 3.12. The best NSI model comes with a $\zeta_o$ value of -0.2 and a $\chi^2_{\nu}$ of 2.96. Including extra data from the Darwin experiment in our analysis refines the range of $\zeta_o$ values from $-0.7$ to $0.002$, down to $-0.5$ to $-0.002$.
These results hint at the possible existence of novel interactions, given that NSI models achieve a comparable or superior fit to the solar neutrino data when contrasted with the prevailing standard model of neutrino flavor oscillation.
\end{abstract}

\keywords{The Sun --- Dark Matter --- Solar neutrino problem ---  Solar neutrinos ---
Neutrino oscillations --- Neutrino telescopes --- Neutrino astronomy}

\maketitle

\section{Introduction} \label{sec:intro}

\medskip\noindent
Neutrinos are widely regarded as one of the most valuable probes for studying the Standard Model (SM) of elementary particles and fundamental interactions, thanks to their unexpected behavior when compared to other elementary particles \citep[e.g.,][]{1987RvMP...59..671B,2004NJPh....6..122M}. This insight has been derived from extensive experimental datasets from detectors around the world. Our knowledge of neutrinos spans many different physical contexts and energy scales, from detecting astrophysical neutrinos with energies ranging from MeV to PeV, to producing them in nuclear reactors and accelerators with energies above MeV and GeV, respectively \citep[e.g.,][]{2023PrPNP.12904019S,2018ARNPS..68..313B}.

\medskip\noindent
Astrophysical neutrinos have been historically at the heart of some of the most compelling challenges to modern physics and  astrophysics. This sphere of exploration includes groundbreaking discoveries, such as the detection of solar neutrinos, as evidenced by \citet{1968PhRvL..20.1205D}, and the identification of neutrino production in remarkable events like Supernova 1987A, as reported by \citet{1987PhRvL..58.1490H}  and \citet{1987PhRvL..58.1494B}.
Additionally, the recent discovery of high-energy neutrinos sourced from distant celestial entities, as chronicled by the  \citet{2018Sci...361..147I}, highlights the substantial advancements unfolding within the specialized field of neutrino astronomy.
A historical perspective on the critical role of astrophysical neutrinos within modern physics can be gleaned from comprehensive reviews, such as those by  \citet{2011neph.book.....Z, 2017FrP.....5...70G,2022arXiv220808050F,2022PTEP.2022lB103N}. These phenomena have collectively designated neutrinos as the ultimate messengers of novel physics extending beyond the Standard Model's boundaries.

\medskip\noindent
 Despite the SM providing the framework for how neutrinos interact with leptons and quarks through weak interactions, many fundamental questions remain unanswered, such as the mechanism for neutrino mass generation or whether neutrinos are Dirac or Majorana particles. For a more detailed account, please refer to the comprehensive reviews by \citet{2007RPPh...70.1757M} and 
\citet{2022PrPNP.12403947A}.
These questions provide solid motivation for thoroughly testing the standard picture of the three-neutrino flavor oscillation \citep[e.g.,][]{2006PhR...429..307L,2008PhR...460....1G}. Specifically, neutrino oscillations over the years have presented compelling evidence for novel physics surpassing the boundaries of the Standard Model, as evidenced by \citet{1998PhRvL..81.1562F} and \citet{2002PhRvL..89a1301A}. Consequently, they function as a highly effective tool for examining the possible presence of novel particles and their interactions. With the increasing sensitivity of neutrino experiments \citep{2023EPJC...83...15A}, it is timely to investigate whether there are any new interactions between neutrinos and matter.

\medskip\noindent
The particle physics community has proposed many alternative neutrino physics models to address these questions, including simple extensions to the SM and models addressing the origin of dark matter, dark energy, and experimental neutrino anomalies \citep[e.g.,][]{2012PhRvD..86k3014G,2013PhRvD..87a3004G,2017JCAP...07..021C,2018JCAP...07..004C,2017JHEP...11..099D}.
These models encompass the introduction of novel particles, including new types of fermions and bosons, such as sterile neutrinos and axionlike particles \citep[e.g.,][]{2018EPJC...78..327L,2020ApJ...905...22L,2019ScPP....6...38H,2021arXiv211010074A,2022PrPNP.12403947A,2023EPJC...83...15A}.

\medskip\noindent
In this article, we delve into the impact of a new quark neutrino interaction on the three neutrino flavor oscillation model \citep{2008PhR...460....1G}, which is predicted by the current standard solar model \citep[e.g.,][]{2020MNRAS.498.1992C,2013MNRAS.435.2109L,1993ApJ...408..347T}. This nonstandard Interaction (NSI) model, developed by \citet{2023PhRvD.107c5007B}, provides a compelling explanation for some of the unsettled experimental data, including the coherent elastic neutrino-nucleus scattering ($CE\nu NS$) experiments \citep{2021arXiv210211981E} and the anomalous magnetic dipole moment of the muon ($g-2)_\mu$ \citep{2021PhRvL.126n1801A}. This model is based on a $U(1)$ gauge symmetry, incorporating a light gauge boson that mixes with the photon \citep[e.g.,][]{2015PhLB..748..311F,2022JHEP...07..138C}.

\medskip\noindent
The coupling of neutrinos with up ($u$-) and down ($d$-) quarks leads to a ratio that nullifies the contribution to the $CE\nu NS$ amplitude, relaxing the constraint on the NSI model with the $CE\nu NS$ experimental measurements \citep{2021arXiv210211981E}. Furthermore, the constraints imposed on the parameter space of this model through experimental and observational bounds lead to a solution that is compatible with the $(g-2)_\mu$ anomaly.

 \medskip\noindent
 Here, we present novel constraints on the NSI model using state-of-the-art solar neutrino data and an up-to-date standard solar model \citep[e.g.,][]{2022arXiv220914832X}. Furthermore, we determine the parameter range that is consistent with solar neutrino experimental measurements and predict potential constraints that could be derived from future neutrino experiments.

 \medskip\noindent 
 The article is organized as follows: Sec.\ref{sec:NeuQuarkNSI} provides a summary of the nonstandard quark-neutrino model used in this work. In Sec.\ref{sec-SNSPeleneut}, we calculate the survival probability of electron neutrinos. Next, Sec.\ref{sec-CGFC} presents the constraints obtained from the standard solar model. Finally, Sec.\ref{sec-Con} provides a summary and draws conclusions.
  
%%%%%%%%%%%%%%%%%%%%%%%%%%%%%%%%%%%%%%%%%%%%%%%%%%%%%%%%%%%%%%%%%%%%%%%%%%%%%
\section{Neutrinos and nonstandard Interaction with Quarks} \label{sec:NeuQuarkNSI}  

 Here, we consider an extension to the standard model of elementary particles and fundamental interactions with a new 
 interaction between active neutrinos and up  and down  quarks \citep[e.g.,][]{2016PhRvD..94e3010F,2018FrP.....6...10T,2021JHEP...01..114C,2019ScPP....6...38H,2023EPJC...83...15A}.
 Accordingly, we consider that our model's Lagrangian density
 ${\cal L}$ corresponds to the sum of the standard model's Lagrangian ${\cal L}_{ST}$ plus a nonstandard Interaction ($NSI$) Lagrangian ${\cal L}_{NSI}$. Hence, 
 \begin{eqnarray}
 {\cal L}={\cal L}_{ST}+{\cal L}_{NSI},
 \end{eqnarray}
where 
${\cal L}_{NSI}$  is the effective Lagrangian that describes the $NSI$ contribution resulting from the neutrino propagation in  matter \citep[e.g.,][]{2018JHEP...08..180E,2019ScPP....6...38H}.
In this study, we focus on an extension of the standard model by a new local group $U_{Z^\prime}(1)$. $Z^\prime$  denotes the gauge boson of the $U_{Z^\prime}(1)$ symmetry group. We also assume that  $Z^\prime$  has a mass $m_{Z^\prime}$  and couples to matter with a coupling constant $g_{Z^\prime}$.  The ${\cal L}_{NSI}$ 
corresponds  now to a NSI vectorlike interaction  \citep{2023PhRvD.107c5007B}, such that $	{\cal L}_{NSI}\equiv
{\cal L}_{Z^\prime}	$,  where 
${\cal L}_{Z^\prime}$ is defined as
\begin{eqnarray}
	{\cal L}_{Z^\prime}	=2\sqrt{2}G_F\epsilon_{\alpha \beta}^{f}
	\left(\bar{\nu}_\alpha \gamma_\mu \frac{1-\gamma_5}{2}\nu_\beta \right)	\left(\bar{f}\gamma^\mu f \right),
	\label{eq:LNSI2}	
\end{eqnarray} 
where  $\alpha$ and $\beta$ refer to neutrino flavors  $e$, $\mu$ and $\tau$; and $f$ and $\bar{f}$ correspond to the fermions or antifermions: 
up  quarks, down  quarks and  electrons. 
The previous Lagrangian [Eq. \ref{eq:LNSI2}] 
 corresponds an NSI model  with an arbitrary ratio of NSI  coupling to the $u$ --- and $d$ --- quarks \citep[e.g.,][]{2015PhLB..748..311F,2016JHEP...07..033F,2020PhLB..80335349F}. 
  Since we are interested in only the contribution of the NSI interaction for the neutrino oscillation experiments, only the vector part contributes to the interaction $\epsilon_{\alpha\beta}^f$. Consequently, the coherent forward scattering of neutrino in the matter is unpolarized \citep[e.g.,][]{2019ScPP....6...38H}.  
 In the case where $|\epsilon_{\alpha \beta}^f|\sim 1$, the contribution of NSI becomes as strong as  the weak interaction.  We notice, in the limit that $\epsilon_{\alpha\beta}^f=0$, we obtain the standard case for which
 ${\cal L}={\cal L}_{ST}$ (${\cal L}_{NSI}=0$). 
 
  \medskip\noindent 
 Here, we describe the propagation of neutrinos through vacuum and matter employing the three-flavor neutrino oscillation model
 \citep[e.g.,][]{1989RvMP...61..937K,
 2013JHEP...09..152G,2020ApJ...905...22L}.
As usual, we follow the standard convention,  $(\nu_e,\nu_\tau,\nu_\mu)$,  $(\nu_1,\nu_2,\nu_3)$ and  $(m_1,m_2,m_3)$ correspond to the neutrino flavors, neutrino mass eigenstates and the associated neutrino masses. Accordingly,  the neutrino evolution equation reads
 \begin{eqnarray}
i  \frac{d\Psi}{dr}={\cal H}_\nu \Psi
=\left({\cal H}_{\rm vac}
+{\cal H}_{\rm mat}\right) \Psi
 \label{eq:NuEvolution}	
 \end{eqnarray}
 where $r$ (distance to the center of the Sun) is the coordinate along the neutrino trajectory,   ${\cal H}_{\nu}$ is the Hamiltonian and
 $\Psi=(\nu_e,\nu_\tau,\nu_\mu)^{T}$.
 Conveniently,  we can decompose this  ${\cal H}_\nu$  in a vacuum and matter components: ${\cal H}_{\rm vac}=\mathbf{U}{\rm M}^2 \mathbf{U}^ \dagger/(2E)$  and  ${\cal H}_{\rm mat}\equiv{\cal V}$, where $E$ is the energy of the neutrino,  ${\rm M}^2={\rm diag}(0,\Delta m_{21}^2,\Delta m_{31}^2)$
 is the neutrino mass matrix,   $\mathbf{U}$ is a unitary matrix describing the mixing of neutrinos in vacuum, ${\cal V}$ is a diagonal matrix of Wolfenstein potentials.  
 $\Delta m^2_{21}$ and    $\Delta m^2_{31}$  are the mass-squared differences between neutrinos of different mass eigenstates, such as  $\Delta m^2_{21}= m_2^2-m_1^2$  and   $\Delta m^2_{31}= m_3^2-m_1^2$. Moreover,  we decompose ${\cal V}$  into two additional components \citep{2023PhRvD.107c5007B}, one related to the standard matter interactions and another one to NSI interactions:
 \begin{eqnarray}
{\cal V}={\cal V}^{SM}+{\cal V}^{NSI},
\label{eq:V}
 \end{eqnarray}
where ${\cal V}^{SM}$ is the standard matter Wolfenstein potential defined as ${\cal V}^{SM}={\rm diag}(V^{SM}_e,0,0)$, and ${\cal V}^{NSI}$ is
the NSI matter Wolfenstein potential defined as
 ${\cal V}^{NSI}= {\rm diag}(0,V^{NSI}_\mu,V^{NSI}_\tau)$. 
Therefore, the nonstandard Interactions matrix, symbolized as ${\cal V}^{NSI}$, is characterized as a diagonal $3\times 3$ matrix, mirroring the structure of the standard Wolfenstein potential denoted as ${\cal V}^{SM}$.
  This process corresponds to a generalisation of  the well-known Mikheyev-Smirnov-Wolfenstein effect \citep[MSW;][]{1978PhRvD..17.2369W,1985YaFiz..42.1441M}.  For the standard Wolfenstein potential for neutrino propagation \citep{2008PhR...460....1G}, we conveniently chose to define it as
  \begin{eqnarray}
  % equation (61) from 2008PhR...460....1G	
 V^{SM}_{e}= \sqrt{2}G_{F}n_e(r), 
 \label{eq:VeMS}
 \end{eqnarray}
 where  $G_F$ is the Fermi constant and $n_e(r)$ is the number density of electrons  inside the Sun.
 
 \medskip\noindent 
 In this study we focus on the NSI model proposed by
 \citet{2023PhRvD.107c5007B}. They  have opted to impose in this NSI model the additional condition: the lepton numbers $L_\mu$  and $L_\tau$, the baryon numbers   $B_i$ with flavor $i$ (such that $i=1,2,3$ corresponding to  the three generations) and any arbitrary real value of $c_o$, fulfil the following rule: 
$L_\mu+L_\tau-c_o(B_1+B_2)-2B_3(1-c_o)$, which accommodates the B meson anomalies observed at LHC \citep{2013PhRvL.111s1801A},  under which the model is anomaly-free \citep{2015PhRvD..91g5006C}. The relationship established earlier shows that if we consider an arbitrary real number, such as $c_o\neq 2/3$, then the $U_{Z^\prime}(1)$ charges of the third generation of quarks will differ from those of the first and second generations. In the model calculated by \citet{2023PhRvD.107c5007B}, the nonstandard Interaction  contribution to the potential, which relates to neutrino propagation in matter, assumes a straightforward form: $V^{NSI}_\mu=V^{NSI}_\tau=V_{Z^\prime}$. Here, $V_{Z^\prime}$ is defined as
	\begin{eqnarray}
		V_{Z^{\prime}}= 2\sqrt{2}G_F n_e(r) \epsilon_{Z\prime}(r),
		\label{eq:VNSI2}
	\end{eqnarray}
	demonstrating the relationship between the NSI potential, the Fermi constant ($G_F$), electron density ($n_e$), and the NSI strength parameter ($\epsilon_{Z\prime}$).

In the previous equation, $\epsilon_{Z\prime}(r)$  estimates the contribution of the NSI Lagrangian. Here,  $\epsilon_{Z\prime}(r)$  is given by
\begin{eqnarray}
 \epsilon_{Z\prime}(r)	= \zeta_{o} \frac{n_n(r)+n_p(r)}{n_e(r)},
	\label{eq:epsilonNSI}	
\end{eqnarray} 
where  $ \zeta_{o}= - c_o g_{Z^\prime}^2/(2\sqrt{2} G_F m_{Z^\prime}^2)$, and  $n_n(r)$ and $n_p(r)$ are the number density of neutrons and protons  or  $u$ --- quarks and $d$ --- quarks 
inside the Sun.  We notice $ \zeta_{o}$, like $c_o$ can be a positive or negative value.  A detailed account of this model is available in \citet{2023PhRvD.107c5007B}, and additional information is available in other related articles \citep[e.g.,][]{2007PhRvD..75k5001F,2021arXiv210403297A}.
% We note that the medium is electrically neutral such that $2/3\, n_u-1/3\,n_d-n_e=0$. 
Furthermore, we will assume that the $Z^\prime$ boson's mass is sufficiently large, and there is no need to consider the size of the medium in the computation of the Welfonstein potentials  \citep{2019JHEP...12..046S}. 

 \medskip\noindent 
 The standard three-flavor neutrino oscillation model features a universal term, denoted as $V_e^{SM}$, that applies to all active neutrino flavors and does not alter the flavor oscillation pattern. This allows us to simplify the model by setting ${\cal V}={\cal V}^{SM}\equiv{\rm diag}(V^{SM}_e,0,0)$
 Now, the inclusion of NSI interaction in the  model alters ${\cal V}$ [see Eq. \ref{eq:V}] by incorporating a new interaction with $u$ --- and $d$ --- quarks, as a consequence 
${\cal V}={\rm diag}(V^{SM}_e,V_{Z^\prime},V_{Z^\prime})$.
Now, if we subtract the common term, $V_{Z^\prime}$ [Eq. \ref{eq:VNSI2}]  to the diagonal matrix ${\cal V}$ \citep[e.g.,][]{2020ApJ...905...22L}, the latter takes the simple form ${\cal V}={\rm diag}(V_{\rm eff},0,0)$ with $V_{\rm eff}\equiv V^{SM}_e-V_{Z^\prime}$ defined as:
 \begin{eqnarray}
V_{\rm eff}= \sqrt{2}G_F\, n_{\rm eff}(r)
	\label{eq:Veff}
\end{eqnarray}
and $	n_{\rm eff}(r)$ is the effective number density given by
 \begin{eqnarray}
	n_{\rm eff}=  n_e(r)
	\left[1-2 \epsilon_{Z\prime}(r)
	\right],
	\label{eq:neff}
\end{eqnarray}
where $\epsilon_{Z\prime}$ is given by Eq.
(\ref{eq:epsilonNSI}).
%%%%%%%%%%%%%%%%%%%%%%%%%%%%%%%%%%%%%%%%%%%%%%%%%%%%%%%%%%%%%%%%%%%%%%%%%%%%%
\section{Solar Neutrinos:  Survival probability of electron neutrinos}
\label{sec-SNSPeleneut}
%%%%%%%%%%%%%%%%%%%%%%%%%%%%%%%%%%%%%%%%%%%%%%%%%%%%%%%%%%%%%%%%%%%%%%%%%%%%%
\begin{figure}[!t] % $m_\phi=10^{-9}\; {\rm eV}$  
\centering 
\includegraphics[scale=0.45]{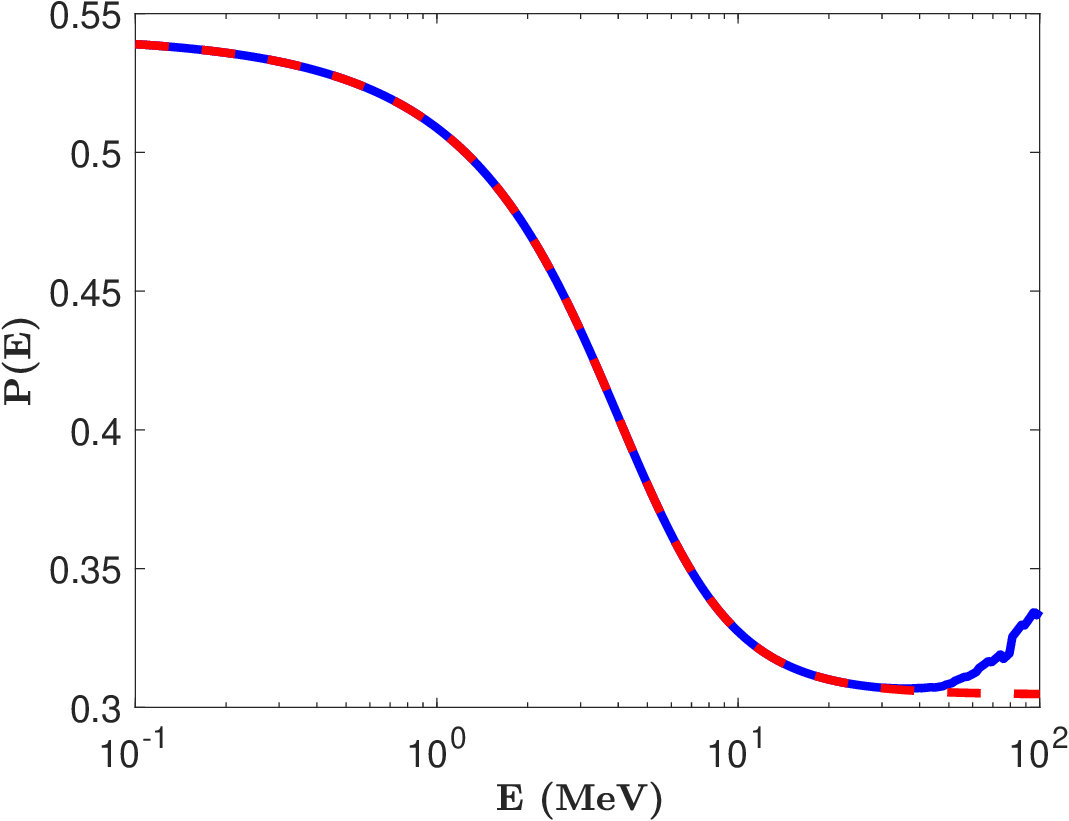}
\caption{
The survival probability of the electron neutrino 
$P_{e}(E)$ [Eqs. \ref{eq:Pee} and \ref{eq:Pee2}]
 computed for the standard model of neutrino flavor oscillations. We use an updated version of the standard solar model for this calculation. See main text for details. 
 We compute  $P_{e}(E)$ including the term $P_{\gamma}$ [Eq. \ref{eq:Pgamma}] in two cases: one where we include the probability jump term $P_{\gamma}\ne 0$ (continuous blue curves)  and a second one for which $P_{\gamma}= 0$ (dashed red curve).  $P_{\gamma}$ is negligible for most of the neutrino energy interval shown, becoming marginally significant for $E\ge 50\;{\rm MeV}$.}
	\label{fig:Peeeff}
\end{figure}

 \medskip\noindent 
We compute the survival probability of electron neutrinos $P_{e}(E)$  of several NSI models with different $\zeta_o$ [Eq. \ref{eq:epsilonNSI}] values and compare them with the data from recent solar neutrino experiments.  
 Several groups have shown that, at a reasonable approximation, the neutrino flavor oscillations are adiabatic  \citep{2004NJPh....6...63B,2017ChPhC..41b3002B,
2021Univ....7..231K}. As such, we can compute a full analytical $P_{e}(E)$  expression that agrees with the current solar neutrino data \citep[e.g.,][]{2013PhRvD..88d5006L}. Moreover, many authors opted to include  a second-order nonadiabatic contribution in $P_{e}(E)$  by  modifying the original adiabatic $P_{e}(E)$  expression
 \citep[e.g.,][]{1986PhRvL..57.1271H,1986PhRvL..57.1275P,2013PhRvD..88d5006L,2013ARA&A..51...21H,2017ChPhC..41b3002B}. 
 The reader can find a detailed discussion about nonadiabatic neutrino flavor oscillations in many articles, among others, the following ones:  \citet{2003RvMP...75..345G,2018arXiv180205781F}.
 
 \medskip\noindent  
 Here, we follow a recent review of particle physics on this topic \citep{2018PhRvD..98c0001T}, specifically in the computation described in the "Neutrino Masses, Mixing, and Oscillations" section \citep[the update of November 2017]{2016ChPhC..40j0001P}. The survival probability of electron neutrinos $P_{e}(E)$   is given by 
 % 2017ChPhC..41b3002B: equations (1-2-3)
 % 2016ChPhC..40j0001P  Equation (14.90)
 \begin{eqnarray}
 	P_{e}(E)\approx\cos^4{(\theta_{13})}P_{e}^{2\nu_e}+\sin^4{(\theta_{13})}
 	\label{eq:Pee}	
 \end{eqnarray}
and 
% 2016ChPhC..40j0001P  Equations (14.85) and (14.86)  
\begin{eqnarray}
	P_{e}^{2\nu_e}(E)=\frac{1}{2}+\left(\frac{1}{2}-P_{\gamma}\right)\cos{(2\theta_{12})} 
	\cos{(2\theta_{m})}.
	\label{eq:Pee2}		
\end{eqnarray}
In the previous expression, $P_{e}^{2\nu_e}(E)$  gives the survival probability of electron neutrinos in the two neutrino flavor model ($\theta_{13}=0$),  $P_{\gamma}$  computes the probability jumps coming from the nonadiabatic correction, and $\theta_{m}=\theta_{m}(r_s)$   is the matter mixing angle \citep{2003NIMPA.503....4D}.  $\theta_{m}$   is evaluated in the neutrino production (source) region located at a distance $r_s$ from the Sun's center \citep[e.g.,][]{1989PhRvD..39.1930K,1995PhRvD..51.4028B}.
The  jump probability $P_\gamma$   reads  
% 2016ChPhC..40j0001P  Equation (14.87)  
\begin{eqnarray}
	P_\gamma=\frac{e^{-\gamma\sin^2{\theta_{12}}}-e^{-\gamma}}{1-e^{-\gamma}} P_{\rm H}
	\label{eq:Pgamma}	
\end{eqnarray}
where $\gamma=2\pi h_\gamma \Delta m_{21}^2/2E$, $h_\gamma$ is the scale height \citep{2000PhLB..490..125G} and
 $P_{\rm H}$ is a regular step function. The  matter mixing angle  \citep{2011PhRvD..83e2002G} $\theta_m$ is  given by
% 2016ChPhC..40j0001P  Equations (14.85) and (14.86)  
\begin{eqnarray}
	\cos(2\theta_{m})=\frac{A_m}{
		\sqrt{A_m^2 +\sin^2{(2\theta_{12})}  }}
	\label{eq:sintheta12}	
\end{eqnarray}
where $A_m$  reads
\begin{eqnarray}
	A_m=\cos{(2\theta_{12})}-{V_m}/{\Delta m^2_{21}}. 
	\label{eq:Am}	
\end{eqnarray}
In the standard case \citep{2004NJPh....6...63B},  it corresponds to
$V_{m}=2V_e^{SM}\cos^2{(\theta_{13})}E$ where $V_e^{SM}$ is given  by  Eq. (\ref{eq:VeMS}).  However, $V_e^{SM}(r)$  in this study will  be replaced by a new effective potential $V_{\rm eff}(r)$ given by Eq. (\ref{eq:Veff}),
with $n_{\rm eff}(r)$ by Eq. (\ref{eq:neff}).

\medskip\noindent 
We remind the reader that we use standard parametrization for the neutrino flavor oscillations: mass square splitting and angle between neutrinos of different flavors \citep[e.g.,][]{2016NuPhB.908..199G}.
Hence, we  adopt the recent values obtained by the data analysis 
of the standard three-neutrino flavor oscillation model obtained by
\citet{2021JHEP...02..071D}.
  Accordingly, for a parametrization with  a normal ordering of neutrino masses the mass-square difference and the mixing angles have the following values \citep[see table 3 of][]{2021JHEP...02..071D}:	
$\Delta m^2_{21}= 7.50^{+0.22}_{-0.20}\times 10^{-5}{\rm eV^2}$, 
$\sin^2{\theta_{12}}=0.318\pm 0.016 $,
and  $\sin^2{\theta_{13}}=0.02250^{+0.00055}_{-0.00078}$.
Similarly  $\Delta m^2_{31}= 2.55^{+0.02}_{-0.03}\times 10^{-3}{\rm eV^2}$ and $\sin^2{\theta_{23}}=0.574\pm 0.014 $.

\medskip\noindent 
The maximum production of neutrinos in the Sun's core occurs in a region between 0.01 and 0.25 solar radius, with neutrino nuclear reactions of the proton-proton chain and carbon-nitrogen-oxygen cycle occurring at different locations \citep[e.g.,][]{2013ApJ...765...14L,2013MNRAS.435.2109L}. These neutrinos produced at various values of $r_s$, when traveling towards the Sun's surface, follow paths of different lengths. Moreover, neutrinos experience varying plasma conditions during their traveling, including a rapid decrease of the electron density from the center towards the surface. In general, we expect that nonadiabatic corrections averaged out and be negligible along the trajectory of the neutrinos, except at the boundaries (layer of rapid potential transition) of the neutrino path, typically around the neutrino production point or at the surface of the Sun.
Therefore, we could expect Eq. (\ref{eq:Pee2})   to be very different when considering such effects.
Nevertheless, this is not the case: \citet{2004NuPhB.702..307D}  analysed in detail the contribution to $P_{e}$
[Eq. \ref{eq:Pee}] coming from nonadiabaticity corrections and variation on the locations of neutrino production, i.e.,  $r_s$, and they found that the impact is minimal. 
Generally,  $P_\gamma=0$   [Eq. \ref{eq:Pgamma}]  corresponds to an adiabatic flavor conversion and $P_\gamma \ne 0$  to a nonadiabatic one. For reference, the conversion is called nonadiabatic only if $P_\gamma \ne 0$ has a non-negligible value.

\medskip\noindent
We notice that inside the Sun, the number densities of electrons, protons, and neutrons vary considerably among the different neutrino paths.
Accordingly,  $n_e(r)$ , $n_p(r)$  and $n_n(r)$  decrease monotonically from the center towards the surface.  As the neutrinos produced in the core propagate towards the surface, a fraction is converted to other flavors. The magnitude of this conversion depends on the neutrino's energy and the coupling constant to electrons, up quarks and down quarks. We remember that in the standard neutrino flavor oscillation model with $\zeta_o=0$, only the $n_e(r)$  contributes to the matter flavor conversion. However, in our NSI model with $\zeta_o\ne 0$, the $n_p(r)$  and $n_n(r)$  also participate in the flavor conversion.

\medskip\noindent
Neutrinos in their path will cross a layer where $A_m=0$ [Eq. \ref{eq:Am}]. This layer is defined by the resonance condition:
\begin{eqnarray}
	V_{m}= \Delta m^2_{21}\cos{(2\theta_{12})}.
	\label{eq:Am0}	
\end{eqnarray}
We compute the effective number density  associated with the resonance condition by matching Eqs.  (\ref{eq:Am}) and (\ref{eq:Am0}). Therefore, the	$n_{\rm eff}$ in the resonance layer reads
\begin{eqnarray}
	n^o_{\rm eff}\equiv	
	n_{\rm eff}(r_o)=	
	\frac{ \Delta m^2_{21}\cos{(2\theta_{12})}}{2\sqrt{2}
		G_F E	 \cos^2{(\theta_{31})}},
	\label{eq:Nres}
\end{eqnarray}
where $r= r_o$ ($\ne h_\gamma$) is defined as the layer where  the resonance condition $n_{\rm eff}(r_o)=	n_{\rm res} (E)$ occurs. 
We observe that in the previous equation, $n^o_{\rm eff}(r)$  corresponds to the quantity defined in Eq. (\ref{eq:neff}). Although in the classic case ($\epsilon_{Z^\prime}=0$), the effective number density is equivalent to the  electronic number density in the resonance layer:  $n_{\rm eff}(r_o)=n^o_{e}(r_o)$. In general,  the adiabatic and nonadiabatic nature of neutrino oscillations depends of the neutrino's energy E and the relative value of the resonance condition of  $n_{\rm res}(E)$ [Eq. \ref{eq:Nres}]. For instance,  if a neutrino of energy E is such that: (i)
$n^o_{\rm eff} (E) \gg n_{\rm eff}$
 neutrinos  oscillate practically as in vacuum, (ii) $ n^o_{\rm eff} \ll n_{\rm eff} (E)$ oscillations as suppressed  in the presence  of matter \citep{2016ChPhC..40j0001P}.
 
 %----------------------------------------
 \begin{figure}[!t] % $m_\phi=10^{-9}\; {\rm eV}$  
 \centering 
 \includegraphics[scale=0.50]{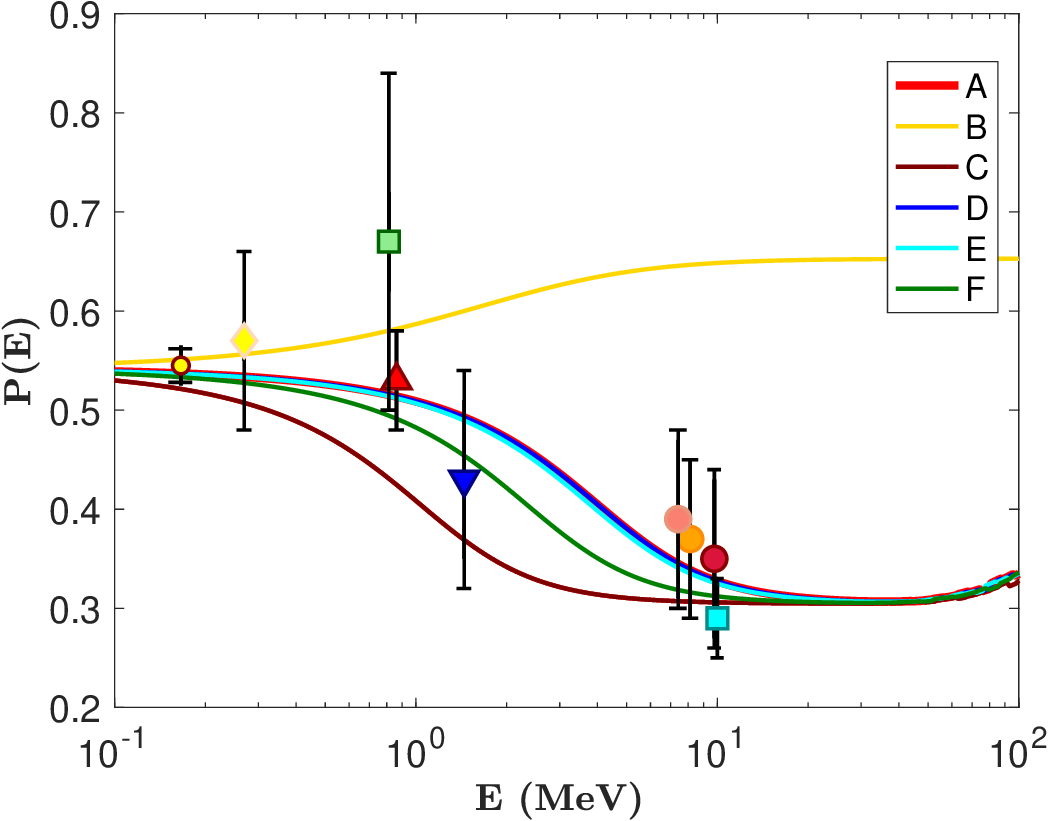}	
 \caption{ 
Survival probability of electron neutrinos in standard and nonstandard Interaction (NSI) neutrino flavor oscillation models with distinct coupling to up and down quarks. Colored continuous curves represent $P_{e}(E)$ [Eq. \ref{eq:Pee}] for various NSI models, accompanied by the corresponding ${\chi_\nu}^2$ values calculated using Eq. (\ref{eq:Chi2nu}). The NSI neutrino models include:
$\zeta_o=2$ {\bf (gold curve, B)}: ${\chi_\nu}^2=111.6$;
$\zeta_o=-2$ {\bf (brown curve, C)}: ${\chi_\nu}^2=5.26$;
$\zeta_o=0.002$ {\bf (blue curve, D)}: ${\chi_\nu}^2=3.13$; and
$\zeta_o=-0.04$ {\bf (cyan curve, E)}: ${\chi_\nu}^2=2.99$.
The {\bf red curve (A)} corresponds to the standard neutrino flavor model with ${\chi_\nu}^2=3.12$, and the {\bf green curve (F)} represents the best-fit NSI model with $\zeta_o=-0.5$ and ${\chi_\nu}^2=2.96$. Data points indicate the measured survival probabilities of electron neutrinos by three solar neutrino detectors (SNO, Super-Kamiokande, and Borexino) using a current standard solar model. For further details regarding the figure and data points, please consult the main text and the referenced sources.}
 	\label{fig:Pe}
 \end{figure}

\medskip\noindent
In our models most of the cases correspond to adiabatic transitions, for which $P_\gamma\approx 0$.  Nevertheless, it is possible to compute the contribution of the non adiabatic component $P_\gamma$ to $P_{e}(E)$ by using Eq. (\ref{eq:Pgamma}) and the following prescription: (i) compute the value of $n^o_{\rm eff}$  (using Eq. \ref{eq:Nres}) for each value of $E$ (with fixed values of $\Delta m_{12}^2$, $\theta_{12}$ and $\theta_{13}$), (ii) calculate the scale height $h_\gamma =|n_{\rm eff}/(d n_{\rm eff}/dr)|_{r_o}$ at the point $r_o$ defined as $n_{\rm eff}(r_o)=n^o_{\rm eff}(E)$, (iii) calculate $P_\gamma$ and $\gamma$ for the value of $h_\gamma$. The  scale-height $h_\gamma$ also reads $h_\gamma =|(d \ln{n_{\rm eff}}/dr)^{-1}|_{r_o}$. 
Conveniently, to properly take into account the nonadiabatic correction into Eqs. (\ref{eq:Pee2}) and (\ref{eq:Pgamma}), we included the step function  $P_{\rm H}$, defined as $P_{\rm H} (V_m - \Delta m^2_{21}   \cos{(2\theta_{12})} )$. This function is one for
$ \Delta m^2_{21}   \cos{(2\theta_{12})}\le V_m $, and is 0 otherwise
\citep[e.g.,][]{2000PhRvD..61j5004C}.
Figure \ref{fig:Peeeff} shows $P_{e} (E)$  for the standard neutrino flavor oscillation model. In any case, in this study we focus on the solar neutrino energy window ($0.1$ up to $20$ MeV), as the $P_\gamma$  contribution for $P_{e}(E)$  is negligible. 

\medskip\noindent 
Numerous studies \citep[e.g.,][]{2013ApJ...765...14L,2017PhRvD..95a5023L} have highlighted that the nuclear reactions occurring in the Sun's core produce a significant amount of electron neutrinos. Due to their extensive mean free path, these neutrinos interact minimally with the solar plasma as they travel towards Earth. During their journey, these particles undergo flavor oscillations (neutrino's energy range spans from $0.1$ to $100$ MeV): lower-energy neutrinos experience flavor transformations due to vacuum flavor oscillations, while high-energy neutrinos participate in additional flavor oscillations, courtesy of the MSW effect or matter flavor oscillations
\citep{1978PhRvD..17.2369W,1985YaFiz..42.1441M}.  This additional oscillation mechanism is significantly influenced by both the origin of the neutrino-emitting nuclear reactions and the energy of the produced neutrinos.

\medskip\noindent 
Here, we will investigate the influence of these revised NSI neutrino models on the flux variation of different neutrino flavors. Specifically, we will consider how these variations are affected by the local alterations in the distributions of protons and neutrons. This new flavor mechanism will affect all electron neutrinos produced in the proton-proton (PP) chain reactions and carbon-nitrogen-oxygen (CNO) cycle
\citep{2013PhRvD..88d5006L,2017PhRvD..95a5023L}.
Therefore, the survival probability of electron neutrinos associated with each nuclear reaction will depend on the location of the neutrino source in the solar interior. A detailed discussion of how the location of solar neutrino sources affects $P_e(E)$ [Eq. \ref {eq:Pee}]  can be found on \citet{2013PhRvD..88d5006L,2017PhRvD..95a5023L}.  The average survival probability of electron neutrinos for each nuclear reaction in the solar interior, i.e., $P_{e,k} $ ($\equiv \langle P_{e} (E)\rangle_k $) is computed 
as
\begin{eqnarray} 
	P_{e,k} (E) = 
	A_k^{-1} \int_0^{R_\odot} P_{e} (E,r)\phi_k (r) 4\pi \rho(r) r^2 dr, 
	\label{eq:Pnuek}
\end{eqnarray}  
where  $A_k$  $(=\int_0^{R_\odot}\phi_i (r) 4 \pi \rho (r) r^ 2  \;dr $ in which $ \phi_k (r) $  is the electron neutrino emission function 
for the $k$ solar nuclear reaction$)$ is a normalization constant, and $k$ corresponds  to the following solar neutrino sources: $pp$, $pep$, $^8B$, $^7Be$, $^{13}N$, $^{15}O$ and $^{17}F$.

\medskip\noindent 
The probability of electron-neutrinos changing flavor is influenced by variables tied to both vacuum and matter oscillations and the intrinsic physics of the Sun's interior. In particular, matter flavor conversion significantly relies on the local plasma conditions. Consequently, the quantity of electron neutrinos detected on Earth for each "$k$" species, as indicated by $\Phi_{\otimes,k}(E)$, diverges markedly from the electron neutrinos generated by each neutrino-producing nuclear reaction, denoted as $\Phi_{\odot,k}(E)$. These quantities are related as follows:
\begin{eqnarray} 
\Phi_{\otimes,k} (E) =P_{e,k} (E)\;\Phi_{\odot,k} (E),
	\label{eq:Phinuek}
\end{eqnarray}  
where $P_{e,k} (E)$ [Eq. \ref{eq:Pnuek}] is the electron-neutrino survival probability of a neutrino of energy $E$. In this study $k$ is equal to $^8B$ or $^7Be$.

%----------------------------------------
\begin{figure}[!t] % $m_\phi=10^{-9}\; {\rm eV}$  
%\centering 
\raggedright	
\includegraphics[scale=0.50]{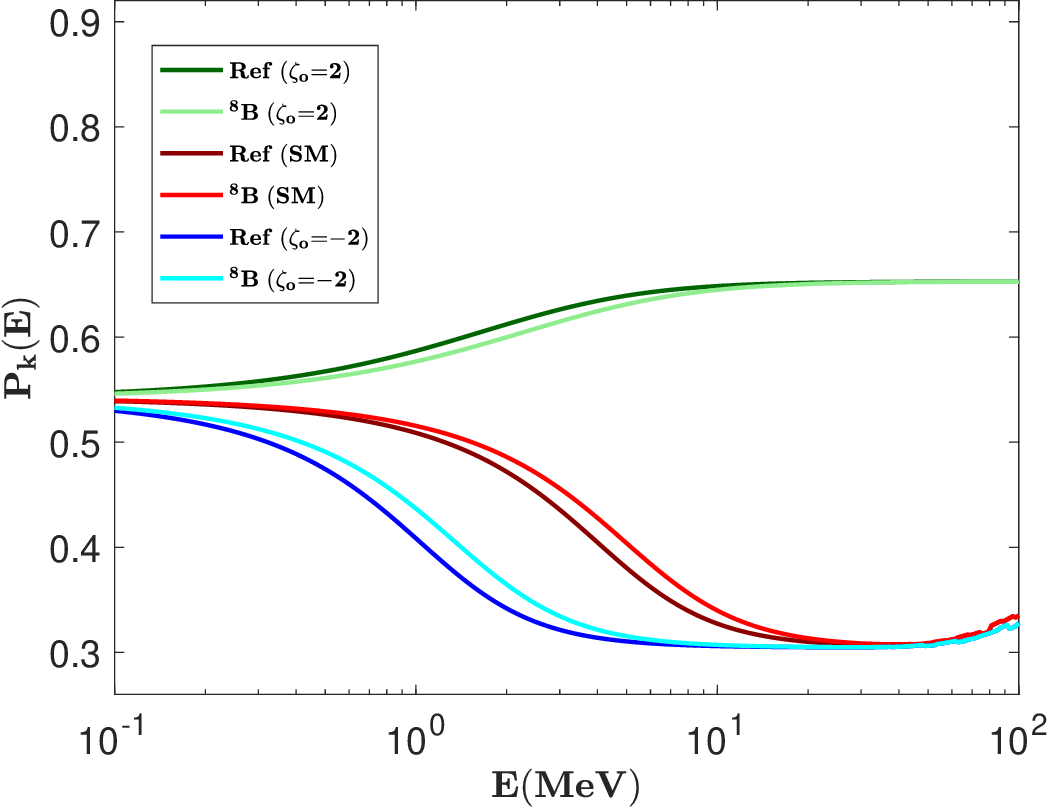}	
\caption{
Survival probability of electron neutrinos: Curves labeled $\mathbf{^8B}$ correspond to neutrinos generated by the $^8B$ nuclear reaction ($\phi_k(r)$), as described in Eq. (\ref{eq:Pnuek}), while the curve labeled as $\mathbf{Ref}$ represents the survival probability of electron neutrinos [Eq. \ref{eq:Pee}] at the Sun's center. The figure presents three distinct sets of $P_{e,k}(E)$ for two NSI models: $\zeta_o=2$ (top set of curves), $\zeta_o=-2$ (lower set of curves), and the standard neutrino flavor model (middle set of curves).}
\label{fig:Pei}
\end{figure}

%----------------------------------------
\begin{figure}[!t] % $m_\phi=10^{-9}\; {\rm eV}$ 
%\centering 	
\raggedright	
\includegraphics[scale=0.47]{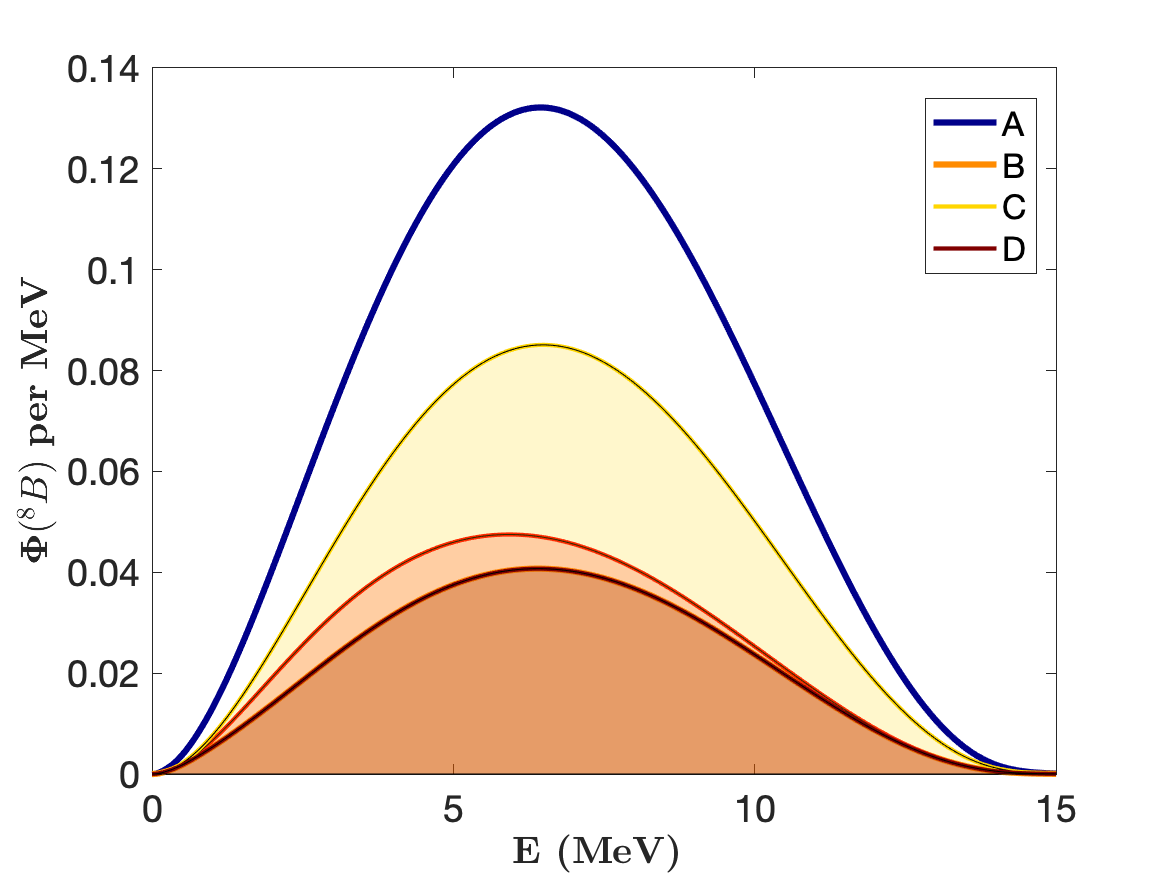}	
\caption{$^8B$ solar neutrino spectrum [refer to Eq. \ref{eq:Phinuek}]: $\Phi_{\otimes}(E)$ represents the electron neutrino energy spectrum of $^8B$ neutrinos for the current Sun, measured on Earth and computed for two NSI neutrino models: $\zeta_o=2$ {\bf (gold area, C)} and $\zeta_o=-2$ {\bf (brown area, D)}, as well as the standard neutrino flavor oscillation model {\bf (orange area, B)}. The {\bf dark blue curve (A)} corresponds to $\Phi_{\odot}(E)$, the neutrino spectrum emitted from the Sun's interior. These neutrino spectra calculations utilize an up-to-date standard solar model.}
\label{fig:8BSpectrum}
\end{figure}

%%%%%%%%%%%%%%%%%%%%%%%%%%%%%%%%%%%%%%%%%%%%%%%%%%%%%%%%%%%%%%%%%%%%%%%%%%%%%%%%%%%%%%%%%%%%%%%% 
\section{Constraints to NSI Neutrino Model} 
\label{sec-CGFC}

\medskip\noindent 
We now turn our attention to the impact of the nonstandard Interactions model on neutrino flavor oscillations, as explored in previous sections. Specifically, we calculate the survival probability of electron neutrinos for varying values of the NSI parameter $\zeta_o$ [as per Eq. \ref{eq:Nres}]. This analysis applies to an updated standard solar model characterized by low metallicity, or 'low-Z.' A comprehensive explanation of the origins of low-Z solar models is presented in the review article of \citet{2013ARA&A..51...21H}. For further exploration of the impact of low metallicity on solar modelling, we refer to the articles by \citet{2009ApJ...705L.123S},
\citet{2017ApJ...835..202V} and	 \citet{2020MNRAS.498.1992C}.

\medskip\noindent 
We obtain the present-day Sun's internal structure using an up-to-date standard solar model that agrees relatively well with current neutrino fluxes and helioseismic datasets. To that end, we use a one-dimensional stellar evolution code that follows the star's evolution from the premain sequence phase until the present-day solar structure: age, luminosity and effective temperature,  $4.57\,{\rm Gyr}$, $3.8418\times 10^{33}\,{\rm erg\,s^{-1}}$, and $5777\,^{\rm o}{\rm K}$, respectively. Moreover, our solar reference model has the following observed abundance ratio at the Sun's surface: $(Z_s/X_s)_\odot=0.01814$, where $Z_s$ and $X_s$ are the metal and hydrogen abundances at the star's surface \citep{1993ApJ...408..347T,2006ApJS..165..400B,2020MNRAS.498.1992C}. The details about the physics of this standard solar model in which we use the AGSS09 (low-Z) solar abundances \citep{2009ARA&A..47..481A} are described in \citet{2013MNRAS.435.2109L}, and \citet{2020MNRAS.498.1992C}.

\medskip\noindent
Figure  \ref{fig:Pe}  compares our predictions with current solar neutrino data. Each data point illustrated herein represents the measured survival probabilities of electron-neutrinos, as captured by three solar neutrino detectors: SNO, Super-Kamiokande, and Borexino.
In detail: Borexino data include measurements from $pp$ reactions (yellow diamond), $^7Be$ reactions (red upward triangle), $pep$ reactions (blue downward triangle), and $^8B$ reactions in the high-energy region (HER), presented in salmon (HER), orange (HER-I), and magenta (HER-II) circles. SNO's $^8B$ measurements are denoted by a cyan square, while the joint KamLAND/SNO $^7Be$ measurements are represented by a green square. Refer to \citet{2018Natur.562..505B,2019PhRvD.100h2004A,2010PhRvD..82c3006B,2011PhRvC..84c5804A,2016PhRvD..94e2010A,2013PhRvC..88b5501A,2008PhRvD..78c2002C} and included references for additional insight into this experimental data.
The lowest neutrino energy data point relates to the anticipated precision of the Darwin experiment in measuring $P_e\pm \Delta P_e$ ($\zeta_o=0$). Here, $\Delta P_e$ has the potential to be as reduced as 0.017, as suggested by \citet{2020arXiv200603114A}.
 Here, we compute $P_{e}$ for several NSI models as given by Eq. (\ref{eq:Pee}). 
 It shows $P_{e}$ for the standard three neutrino flavor model (continuous red curve) and different NSI models  (other continuous colored curves). 
Only a restricted set of NSI models with relatively low $\zeta_o$ agree with all the neutrino data. Notably, the NSI models with lower $\zeta_o$ have an explicit agreement with the $^8B$ measurements for neutrino energies just below $10\, MeV$ (as depicted in Fig. \ref{fig:Pe}).

\medskip\noindent

For illustration, we present a selection of NSI models that significantly diverges from the standard flavor oscillation model in their impact on $P_{e}$. The degree of effect in these NSI models depends on the value of $\zeta_o$, the location of neutrino emission, and the energy spectrum of neutrinos from each nuclear reaction. We illustrate this impact in Figs. \ref{fig:Pei} and \ref{fig:8BSpectrum}, demonstrating how the parameter $\zeta_o$ influences neutrino flavor oscillation [refer to Eq. \ref{eq:Pnuek}] and modulates the $^8B$ spectrum [see Eq. \ref{eq:Phinuek}].
To exemplify the influence of the neutrino source location on $P_{e}$, Fig.~\ref{fig:Pei} displays curves based on the presumption that neutrinos originate from the Sun's center, indicated as "Ref". These curves are then juxtaposed with those derived from neutrinos generated by the $^8B$ nuclear reaction for a variety of $\zeta_o$ values.	

\medskip\noindent
To enhance the robustness of our analysis, we opt to calculate a chi-squared-like test ($\chi_\nu^2$ -- test). This test leverages the inherent reliance of $P_{e}$ on the solar background structure.  Therefore, we define this chi-squared-like test as follows:
\begin{eqnarray} 
	\chi^2_{\nu}= \sum_{i,k}
	\left(\frac{P_{e,k}^{obs}(E_{i})-P_{e,k}^{th}(E_{i})}{\sigma_{obs(E_{i})}} \right)^2.
	\label{eq:Chi2nu}
\end{eqnarray}  

\medskip\noindent
This function compares our theoretical predictions with the empirical data collected by various neutrino experiments, evaluated at different energy values, E, used to calculate the survival probability function $P_{e,k}(E)$, as defined in Eq.  (\ref{eq:Pnuek}).  
Here, the subscript "obs" and "th" signify the observed and theoretical values, respectively, at the neutrino energy $E_i$. The subscript $i$ points to specific experimental measurements [refer to Fig. \ref{fig:Pe}], and $k$ corresponds to the source of solar neutrino [see Eq. \ref{eq:Pnuek}]. The term $\sigma_{obs}(E_{i})$ represents the error in measurement $i$. The data points, $P_{e,k}^{obs}(E_{i})$, are measurements derived from solar neutrino experiments, as cited in  \citet{2018Natur.562..505B,2019PhRvD.100h2004A,2010PhRvD..82c3006B,2011PhRvC..84c5804A,2016PhRvD..94e2010A,2013PhRvC..88b5501A,2008PhRvD..78c2002C}. 
Fig. \ref{fig:Pe} presents the experimental data points, $P_{e,k}^{obs}(E_{i})$, juxtaposed with the curves of select NSI models. The corresponding $\chi^2_{\nu}$ values for these models are explicitly listed in the figure's caption.
In the $\chi_\nu^2$ test, as described by Eq. (\ref{eq:Chi2nu}), the standard neutrino flavor model yields a $\chi_\nu^2$ value of 3.12. 

\begin{figure} % $m_\phi=10^{-9}\; {\rm eV}$  
%$$\qquad$$	
\centering 
\includegraphics[scale=0.45]{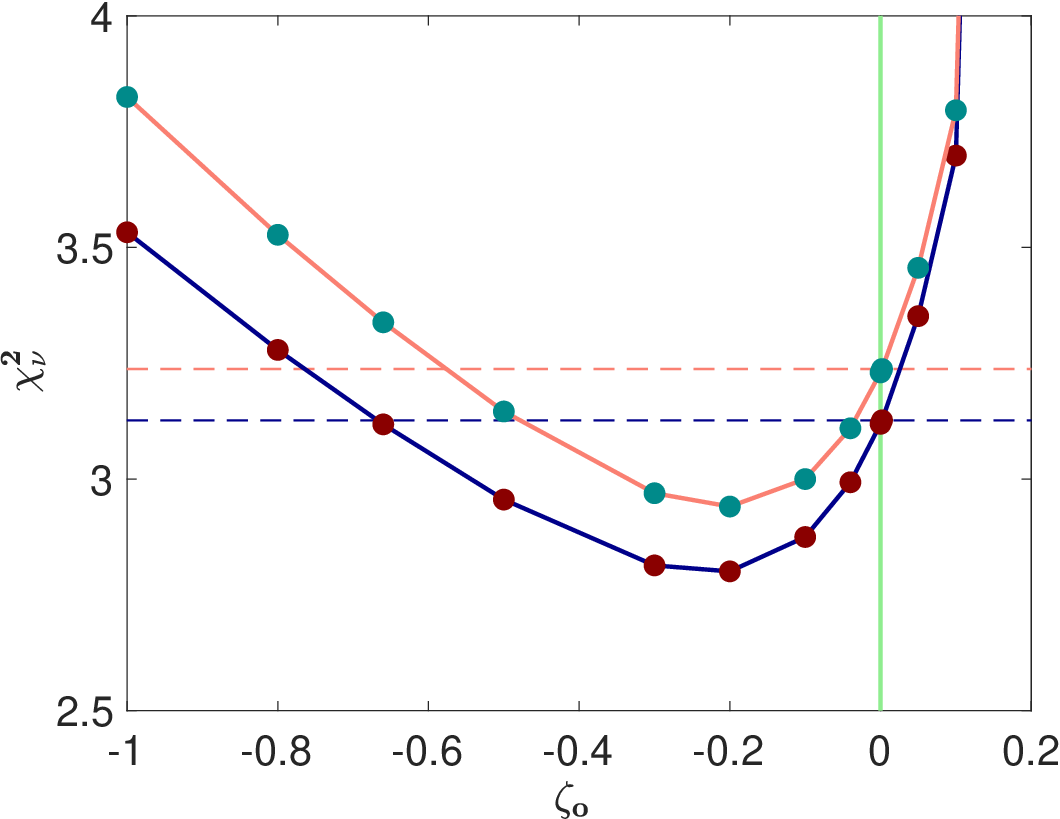}	
\caption{
Values of the $\chi^2_\nu$-test plotted against the coupling constant $\zeta_o$ for various NSI neutrino models. The red circles, interconnected by a blue line, represent varying $\chi^2_\nu$ values corresponding to these NSI models, while the green vertical line signifies the $\chi^2_\nu$ of the standard neutrino model. The green circles, linked with the salmon line, correspond to the same NSI model, including a data point from the Darwin experiment. In this calculation, we assign the Darwin data point a value corresponding to $P(E)$, assuming $\zeta_o=0$. The horizontal dashed lines guide delineating the range between two sets of NSI models - those without the Darwin data point (represented by the blue line) and those incorporating the Darwin data point (depicted by the salmon line).
}
\label{fig:Chi2}
\end{figure}

For comparison, when the $\zeta_o$ values are at -2 and 2, the corresponding $\chi_\nu^2$ values are 5.26 and 111.6, respectively. Our study reveals that a $\chi_\nu^2$ value of 3.12 or less is achieved when $\zeta_o$ lies between -0.7 and 0.002. This result is visually demonstrated in Fig. \ref{fig:Chi2} with a dashed horizontal line intersecting the blue curve, which connects the series of red circles at the points -0.7 and 0.002. According to this preliminary analysis, an NSI neutrino model with $\zeta_o=-0.2$ yields a $\chi_\nu^2$ value of 2.96, suggesting a better fit to the solar neutrino data than the standard neutrino flavor model.

\medskip\noindent
$\qquad$
%%%%%%%%%%%%%%%%%%%%%%%%%%%%%%%%%%%%%%%%%%%%%%%%%%%%%%%%%%%%%%%%%%%%%%%%%%%%%%%%%%%%%%%%%%%%%%%% 
 \section{Conclusion} 
\label{sec-Con}

\medskip\noindent
Currently, a new class of models based on flavor gauge symmetries with a lighter gauge boson is being proposed in the literature to resolve some of the current particle anomalies in the standard model of physics. These new interactions lead to nonstandard neutral current interactions between neutrinos and quarks. Specifically, we focus on studying and testing an NSI model proposed by \citet{2023PhRvD.107c5007B} that incorporates a new U(1) gauge symmetry through a light gauge boson $Z^\prime$, which mixes with the photon. The interaction leads to a neutral current between active neutrinos and matter fields, with an arbitrary coupling to the up and down quarks. This model has some intriguing features, as it relaxes the bound on the coherent elastic neutrino-nucleus scattering experiments and fits the measured value of the anomalous magnetic dipole moment of the muon.

\medskip\noindent
In this paper, we analyze the impact of the NSI model proposed by \citet{2023PhRvD.107c5007B} on neutrino flavor oscillations, using an up-to-date standard solar model that is in good agreement with helioseismology and neutrino flux datasets.
Specifically, we examine the impact of this nonstandard Interaction model on the survival probability of electron neutrinos, with a focus on the PP-chain nuclear reactions taking place in the Sun's core. Our results show that the shapes of the neutrino spectra vary with the location of the nuclear reactions in the core, depending on the algebraic value of $\zeta_o$. The effect is particularly visible in the $^8B$ neutrino spectrum.

\medskip\noindent
We find that the NSI models with $-0.7 \le \zeta_o \le 0.002$ fit the solar neutrino data equal or better than the standard neutrino flavor model. The best NSI model corresponds to $\zeta_o=-0.2$. From Eq. (\ref{eq:epsilonNSI}), we can derive a relationship between the mass of the $Z^\prime$ boson $m_{Z^\prime}$, the gauge coupling $g_{Z^\prime}$, and the quark charge $c_o$: $\zeta_{o}= - c_o g_{Z^\prime}^2/(2\sqrt{2} G_F m_{Z^\prime}^2)=-0.2$.

\medskip\noindent
In essence, our research underscores the significance of neutrino oscillation analyses in assessing NSI models. Our findings reveal the potential of these neutrino models to refine the parameters of NSI models. This methodology provides a robust and independent means to confirm this class of NSI models, especially as they address certain existing experimental data anomalies, such as those observed in coherent elastic neutrino-nucleus scattering experiments and in measurements of the muon's anomalous magnetic dipole moment.

\medskip\noindent
In the future, the validation or exclusion of such a class of NSI models can be achieved more efficiently with new solar neutrino detectors that can obtain much more accurate measurements \citep[e.g.,][]{2019PhRvL.123m1803C,2020JHEP...09..106D,2022EPJC...82..116G}. 
For instance, the Darwin experiment \citep{2020arXiv200603114A} is set to generate data that can better calculate the survival rate of low-energy electron neutrinos [see Fig. \ref{fig:Pe}]. The figure shows that by factoring in the predicted precision from Darwin and presuming the $P(E)$ value to be standard at $E=0.150\; MeV$ (with $\zeta_o=0.0$), we anticipate a $P_e\pm \Delta P_e$ where $\Delta P_e= 0.017$. This additional data point from Darwin, when included in the $\chi^2$ analysis, narrows down the set of NSI models that perform equal or better than the standard case in terms of $\chi^2$. Specifically, it shifts the $\zeta_o$ interval from $-0.7$ to $0.002$ to a tighter range of $-0.5$ to $-0.002$. Furthermore, the addition of this data point also decreases the $\chi^2/{\rm d.o.f.}$ value. For reference, in Fig. \ref{fig:Chi2}, the models with a ${\rm d.o.f.}$ of 7 display $\chi_\nu^2/{\rm d.o.f.}$ values that vary from $0.50$ to $0.53$ within the $\zeta_o$ range of -1 to 0.2, and hit a local minimum of $\chi_\nu^2/{\rm d.o.f.}=0.4$ at $\zeta_o=-0.2$. Adding one more data point increases the d.o.f to 8 and adjusts the $\chi_\nu^2/{\rm d.o.f.}$ range to $0.48$ to $0.47$. The local minimum remains at $\zeta_o=-0.2$, but its value reduces to $\chi_\nu^2/{\rm d.o.f.}=0.37$.

\medskip\noindent
This work emphasizes the significance of NSI models in defining the fundamental properties of particles and their interactions, driving theoretical progress in this research field. As research in experimental neutrino physics continues to advance at a rapid pace, studies of this nature will be critical for comprehensive analysis of neutrino properties \citep{2022arXiv221113450B}. We anticipate that the innovative approach outlined in this paper will offer a fresh perspective for exploring new particle physics interactions using the standard solar model combined with a comprehensive analysis of neutrino flavor oscillation experimental data.

%%%%%%%%%%%%%%%%%%%%%%%%%%%%%%%%%%%%%%%%%%%%%%%%%%%%%%%%%%%%%%%%%%%%%%%%%%%%%
%\acknowledgments
\section*{Acknowledgments}

\medskip\noindent
The author thanks the anonymous referee for the invaluable input which significantly enhanced the quality of the manuscript. I.L. would like to express gratitude to the Funda\c c\~ao para a Ci\^encia e Tecnologia (FCT), Portugal, for providing financial support to the Center for Astrophysics and Gravitation (CENTRA/IST/ULisboa) through Grant Project No. UIDB/00099/2020 and Grant No. PTDC/FIS-AST/28920/2017.


%\bibliographystyle{mnras}
%\bibliographystyle{aasjournal}
%\bibliography{artmnrasbib} % if your bibtex file is called
%\bibliography{artnus23IL}
% example.bib
%\bibliographystyle{aasjournal}

%\input{artnusDM22ILopeslib}
%\bibliographystyle{aasjournal}

%%%%%%%%%%%%%%%%%%%%%%%%%%%%%%%%%%%%%%%%%%%%%%%%%%%%%%%%%%%%%%%%%%%%%%%%%%%%%
 
\end{document}